\documentclass[12pt,draftcls,onecolumn]{IEEEtran}
\usepackage{cite}
\usepackage[font=small]{caption}
\usepackage{ifpdf}
\usepackage{authblk}
\usepackage{commath}
\usepackage{cuted} 

\usepackage{bigints}
\newif\ifCLASSOPTIONonecolumn       \CLASSOPTIONonecolumnfalse
\newif\ifCLASSOPTIONtwocolumn       \CLASSOPTIONtwocolumntrue
\ifCLASSINFOpdf
   \usepackage[pdftex]{graphicx}
 \else
    \usepackage[dvips]{graphicx}
  \fi
\usepackage{subcaption}
\usepackage{amsmath}
\usepackage{dsfont}
\usepackage{bbold}
\usepackage{amssymb}
\usepackage{algorithmic}
\usepackage{array}
\usepackage{stix}
\usepackage{stfloats}
\fnbelowfloat
\usepackage{bigints}
\begin{document}
%
\title{Outage Analysis of Cooperative NOMA Using Maximum Ratio Combining at Intersections \footnote{This paper has been presented at the wireless and mobile computing, networking and communications (WiMob) 2019, Barcelona, Spain, October 2019 \cite{WiMob}.}}
%
%
%
\author[1 2]{Baha Eddine Youcef~Belmekki}
\author[1]{Abdelkrim ~Hamza}
\author[2]{Beno\^it~Escrig}
\affil[1]{LISIC Laboratory, Electronic and Computer Faculty, USTHB, Algiers, Algeria,}
\affil[ ]{email: $\{$bbelmekki, ahamza$\}$@usthb.dz}
\affil[2]{University of Toulouse, IRIT Laboratory, School
of ENSEEIHT, Institut National Polytechnique de Toulouse, France, e-mail: $\{$bahaeddine.belmekki, benoit.escrig$\}$@enseeiht.fr}
\affil[ ]{}
\setcounter{Maxaffil}{0}
\renewcommand\Affilfont{\small}
\markboth{ }%
{Shell \MakeLowercase{\textit{et al.}}: Bare Demo of IEEEtran.cls for IEEE Journals}
\maketitle

\IEEEpeerreviewmaketitle
\begin{abstract}
The paper investigates the improvement of using maximum ratio combining (MRC) in cooperative vehicular communications (VCs) transmission schemes considering non-orthogonal multiple access scheme (NOMA) at intersections. The transmission occurs between a source and two destination nodes with a help of a relay. The transmission is subject to interference originated from vehicles that are located on the roads. Closed form outage probability expressions are obtained.
We compare the performance of MRC cooperative NOMA with a classical cooperative NOMA, and show that implementing MRC in cooperative NOMA transmission offers a significant improvement over the classical cooperative NOMA in terms of outage probability. We also compare the performance of MRC cooperative NOMA with MRC cooperative orthogonal multiple access (OMA),  and we show that NOMA has a better performance than OMA. Finally, we show that the outage probability increases when the nodes come closer to the intersection, and that using MRC considering NOMA improves the performance in this context. The analysis is verified with Monte Carlo simulations. 
\end{abstract}


\section{Introduction}
\subsection{Motivation}
Road traffic safety is a major issue, and more particularly at intersections since $50\%$ of accidents occurs at
intersections \cite{traficsafety}. Vehicular communications (VCs) offer several applications for accident prevention, or alerting vehicles when accidents happen in their vicinity. Thus, high reliability and low latency communications are required in safety-based vehicular communications. To increase the data rate and spectral efficiency \cite{ding2017application} in the fifth generation (5G) of communication systems, non-orthogonal multiple access (NOMA) is an appropriate candidate as a multiple access scheme. Unlike orthogonal multiple access (OMA), NOMA allows multiple users to share the same resource with different power allocation levels.

\subsection{Related Works}
NOMA is an efficient multiple access technique for spectrum use. It has been shown that NOMA outperforms OMA \cite{saito2013non,dai2015non,islam2017power,ding2014performance,mobini2017full}.  However, few research investigates the effect of co-channel interference and their impact on the performance considering direct transmission \cite{ali2018analyzing,zhang2016stochastic,tabassum2017modeling}, and cooperative transmission \cite{liu2017non}.

Regarding VCs, several works investigate the effect of interference considering OMA in highway scenarios \cite{tassi2017modeling}. As for intersection scenarios, the performance in terms of success probability are derivated \cite{steinmetz2015stochastic,abdulla2016vehicle}. The performance of vehicle to vehicle (V2V) communications are evaluated for multiple intersections scheme in \cite{jeyaraj2017reliability}. In \cite{kimura2017theoretical}, the authors derive the outage probability of a V2V communications with power control strategy. In \cite{article}, the authors investigate the impact of a line of sight and non line of sight transmissions at intersections considering Nakagami-$m$ fading channels.
 The authors in \cite{belmekki2018performance} study the effect of mobility of vehicular communications at road junctions.
In \cite{WCNC,J3,VTC,J4}, the authors respectively study the impact of non-orthogonal multiple access, and cooperative non-orthogonal multiple access with NOMA at intersections. The authors further extended their work to millimeter wave vehicular networks using NOMA in \cite{NoMa,NoMa2}.

Following this line of research, we study the performance of vehicular communications at intersections in the presence of interference considering cooperative NOMA transmissions using maximum ratio combining (MRC).

\subsection{Contributions}
The  contributions of this paper are  as follows:
\begin{itemize}
\item We analyze the performance and the improvement of using MRC in cooperative VCs transmission schemes considering NOMA at intersections in terms of outage probability. Closed form outage probability expressions are obtained.

\item We compare the performance of MRC cooperative NOMA with a classical cooperative NOMA, and show that implementing MRC in cooperative NOMA transmission offers a significant improvement over the classical cooperative NOMA in terms of outage probability.

\item We also compare the performance of MRC cooperative NOMA with MRC cooperative OMA,  and we show that NOMA has a better performance than OMA. 

\item Finally, we show that the outage probability increases when the nodes come closer to the intersection, and that using MRC considering NOMA improves significantly the performance in this context. 

\item All the theoretical results are verified with Monte Carlo simulations.
\end{itemize}

\subsection{Organization}

The rest of this paper is organized as follows. Section II presents the system model. In Section III, NOMA outage behavior is investigated. The Laplace transform expressions are presented in Section IV. Simulations and discussions are in Section V. Finally, we conclude the paper in Section VI.

\section{System Model}
\begin{figure}[]
\centering
\includegraphics[scale=0.65]{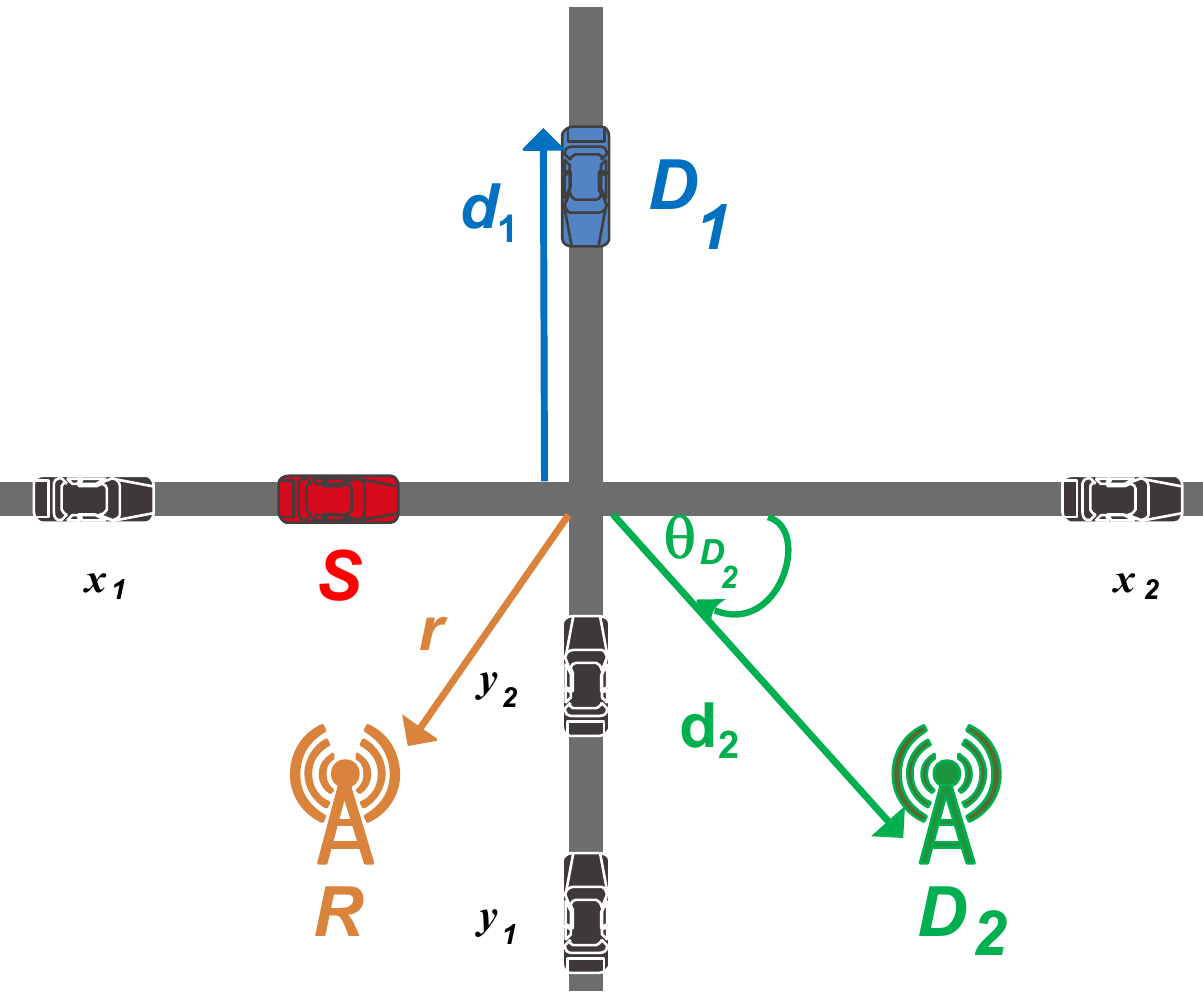}
\caption{Cooperative NOMA system model for vehicular communications involving two destination nodes and a relay node. For this example, $S$ is a vehicle, $R$ is an infrastructure, $D_1$ is a vehicle, and $D_2$ is an infrastructure. }
\label{Fig1}
\end{figure}

In this paper, we consider a cooperative NOMA transmission between a source, denoted $S$, and two destinations, denoted $D_1$ and $D_2$, with the help of a relay, denoted $R$. The set $\{S, R, D_1, D_2\}$ denotes the nodes and their locations as depicted in Fig.\ref{Fig1}.

We consider an intersection scenario involving two perpendicular roads, an horizontal road denoted by $X$, and a vertical road denoted by $Y$.
In this paper, we consider both V2V and V2I communications\footnote{The Doppler shift and time-varying effect of V2V and V2I channel is beyond the scope of this paper}, hence, any node of the set $\lbrace{S, R, D_1, D_2}\rbrace$ can be on the road or outside the roads. We denote by  $M$ the receiving node, and by $m$ the distance between the node $M$ and the intersection, where $M \in \{R,D_1,D_2\}$ and $m \in \{r,d_1,d_2\}$, as shown in Fig.\ref{Fig1}. Note that the intersection is the point where the $X$ road and the $Y$ road intersect.

The set $\lbrace{S, R, D_1, D_2}\rbrace$ is subject to interference that are originated from vehicles located on the roads. The set of interfering vehicles located on the $X$ road, denoted by $\Phi_{X}$ (resp. on the $Y$ road, denoted by $\Phi_{Y}$) are modeled as a One-Dimensional Homogeneous Poisson Point Process (1D-HPPP), that is, $\Phi_{X}\sim\textrm{1D-HPPP}(\lambda_{X},x)$ (resp.$\Phi_{Y}$ $\sim \textrm{1D-HPPP}(\lambda_{Y},y)$), where $x$ and $\lambda_{X}$ (resp. $y$ and $\lambda_{Y}$) are the position of interferer vehicles and their intensity on the $X$ road (resp. $Y$ road). The notation $x$ and $y$ denotes both the interferer vehicles and their locations. We consider slotted ALOHA protocol with parameter $p$, i.e., every node accesses the medium with a probability $p$. We denote by $l_{ab}$ the path loss between the nodes $a$ and $b$, where $l_{ab}= r_{ab}^{-\alpha}$, $ r_{ab}$ is the Euclidean distance between the node $a$ and $ b$, i.e., $ r_{ab}=\Vert a- b\Vert$, and $\alpha$ is the path loss exponent.

We use a Decode and Forward (DF) decoding strategy, i.e., $R$ decodes the message, re-encodes it, then forwards it to $D_1$ and $D_2$. We also use a half-duplex transmission in which a transmission occurs during two phases. Each phase lasts one time–slot. 
We consider using MRC at the destination nodes, hence, during the first phase, $S$ broadcasts the message, and the receiving nodes $R$, $D_1$ and $D_2$ try to decode it, that is, ($S \rightarrow R$, $S \rightarrow D_1$, and $S \rightarrow D_2$). During the second phase, $R$ broadcasts the message to $D_1$ and $D_2$ ($R \rightarrow D_1$ and $R \rightarrow D_2$). Then $D_1$ and $D_2$ add the power received in the first phase from $S$ and the power received from $R$ during the second phase to decode the message.

Several works in NOMA order the receiving nodes by their channel states (see \cite{ding2014performance,ding2015cooperative} and references therein). However, it has been shown in \cite{ding2016relay,ding2016mimo}, that it is a more realistic assumption to order the receiving nodes according to their quality of service (QoS) priorities. We consider the case when, node $D_1$ needs a low data rate but has to be served immediately, whereas node $D_2$ require a higher data rate but can be served later. For instance $D_1$ can be a vehicle that needs to receive safety data information about an accident in its surrounding, whereas $D_2$ can be a user that accesses his/her internet connection. We consider an interference limited scenario, that is, the power of noise is neglected. Without loss of generality, we assume  that all nodes transmit with a unit power. 
The signal transmitted by $S$, denoted $ \chi_{S}$ is a mixture of the message intended to $D_1$ and $D_2$. This can be expressed as
\begin{equation}\label{eq.1}
 \chi_{S}=\sqrt{a_1}\chi_{D1}+\sqrt{a_2}\chi_{D2}, \nonumber
 \end{equation}
where $a_i$ is the power coefficients allocated to $D_i$, and $\chi_{Di}$ is the message intended to $D_i$, where $i \in \{1,2\}$. Since $D_1$ has higher power than $D_2$, that is $a_1 \ge a_2$, then $D_1$ comes first in the decoding order. Note that, $a_1+a_2=1$.\\
 The signal received at $R$ and $D_i$ during the first time slot are expressed as
 \begin{equation}
   \mathcal{Y}_{R}=h_{SR}\sqrt{l_{SR}}\:\chi_{S}+ 
 \sum_{x\in \Phi_{X_{R}}}h_{Rx}\sqrt{l_{Rx}}\:\chi_{x} 
 +\sum_{y\in \Phi_{Y_{R}}}h_{Ry}\sqrt{l_{Ry}}\:\chi_{y}, \nonumber
 \end{equation}
 and 
 \begin{equation}
   \mathcal{Y}_{D_i}=h_{SD_i}\sqrt{l_{SD_i}}\:\chi_{S}+ 
 \sum_{x\in \Phi_{X_{D_i}}}h_{D_{ix}}\sqrt{l_{D_{ix}}}\:\chi_{x} 
 +\sum_{y\in \Phi_{Y_{D_i}}}h_{D_{iy}}\sqrt{l_{D_{iy}}}\:\chi_{y}. \nonumber
 \end{equation}
 The signal received at $D_i$ during the second time slot is expressed as
 \begin{equation}
   \mathcal{Y}_{D_i}=h_{RD_i}\sqrt{l_{RD_i}}\:\chi_{R}+ 
 \sum_{x\in \Phi_{X_{D_i}}}h_{D_{ix}}\sqrt{l_{D_{ix}}}\:\chi_{x} 
 +\sum_{y\in \Phi_{Y_{D_i}}}h_{D_{iy}}\sqrt{l_{D_{iy}}}\:\chi_{y}, \nonumber
 \end{equation}
where $\mathcal{Y}_{Di}$ is the signal received by $D_i$.
The messages transmitted by the interfere node $x$ and $y$, are denoted respectively by $ \chi_x$ and $\chi_y $, $h_{ab}$ denotes the fading coefficient between node $a$ and $b$, and it is modeled as $\mathcal{CN}(0,1)$. The power fading coefficient between the node $a$ and $b$, denoted $|h_{ab}|^2$, follows an exponential distribution with unit mean.
The aggregate interference is defined as 
\begin{eqnarray}\label{eq.5}
I_{X_{M}}=\sum_{x\in \Phi_{X_{M}}}\vert h_{Mx}\vert^{2}l_{Mx}   \\ 
I_{Y_{M}}=\sum_{y\in \Phi_{Y_{M}}}\vert h_{My}\vert^{2}l_{My} , 
\end{eqnarray}
where $I_{X_{M}} $ denotes the aggregate interference from the $X$ road at $M$, $I_{Y_{M}}$ denotes the aggregate interference from the $Y$ road at $M$, $\Phi_{X_{M}}$ denotes the set of the interferers from the $X$ road at $M$, and $\Phi_{Y_{M}}$ denotes the set of the interferers from the $Y$ road at $M$.

\section{NOMA Outage Behavior}
\subsection{Outage Events}
According to successive interference cancellation (SIC) \cite{hasna2003performance}, $D_1$ is decoded first since it has the higher power allocation, and $D_2$ message is considered as interference. The outage event at $R$ to not decode $D_1$, denoted  $\mathcal{A}_{R_1}(\Theta_1)$, is defined as
\begin{equation}\label{eq.6}
\mathcal{A}_{R_1}(\Theta_1)\triangleq \frac{\vert h_{SR}\vert^{2}l_{SR}\,a_1}{\vert h_{SR}\vert^{2}l_{SR}a_2+I_{X_{R}}+I_{Y_{R}}} < \Theta_1,
\end{equation}
where $\Theta_1=2^{2\mathcal{R}_1}-1$, and $\mathcal{R}_1$ is the target data rate of $D_1$.

Since $D_2$ has a lower power allocation, $R$ has to decode $D_1$ message, then decode $D_2$ message. The outage event at $R$ to not decode $D_2$ message, denoted $\mathcal{A}_{R_2}(\Theta_2)$, is defined as \footnote{Perfect SIC is considered in this work, that is, no fraction of power remains after the SIC process.}

\begin{equation}\label{eq.7}
\mathcal{A}_{R_2}(\Theta_2)\triangleq\frac{\vert h_{SR}\vert^{2}l_{SR}\,a_2}{I_{X_{R}}+I_{Y_{R}}}< \Theta_2,
\end{equation}
where $\Theta_2=2^{2\mathcal{R}_2}-1$, and $\mathcal{R}_2$ is the target data rate of $D_2$.

Similarly, the outage event at $D_1$ to not decode its intended message in the first phase ($S \rightarrow D_1$), denoted $\mathcal{B}_{D_1}(\Theta_1)$, is given by

\begin{equation}\label{eq.8}
\mathcal{B}_{D_1}(\Theta_1)\triangleq\frac{\vert h_{SD_1}\vert^{2}l_{SD_1}\,a_1}{\vert h_{SD_1}\vert^{2}l_{SD_1}a_2+I_{X_{D_1}}+I_{Y_{D_1}}} < \Theta_1.
\end{equation}
Finally, in order for $D_2$ to decode its intended message, it has to decode $D_1$ message. The outage event at $D_2$ to not decode $D_1$ message in the first phase ($S \rightarrow D_2$), denoted $\mathcal{B}_{D_{2-1}}(\Theta_1)$, and the outage event at $D_2$ to not decode its intended message, denoted $\mathcal{B}_{D_{2-2}}(\Theta_2)$, are respectively given by
\begin{equation}\label{eq.9}
\mathcal{B}_{D_{2-1}}(\Theta_1)\triangleq\frac{\vert h_{SD_2}\vert^{2}l_{SD_2}\,a_1}{\vert h_{SD_2}\vert^{2}l_{SD_2}a_2+I_{X_{D_2}}+I_{Y_{D_2}}}< \Theta_1,
\end{equation}
and
\begin{equation}\label{eq.10}
\mathcal{B}_{D_{2-2}}(\Theta_2)\triangleq\frac{\vert h_{SD_2}\vert^{2}l_{SD_2}\,a_2}{I_{X_{D_2}}+I_{Y_{D_2}}}< \Theta_2.
\end{equation}
During the second phase, $D_1$ adds the power received from $S$ and from $R$. Hence, the outage event at $D_1$ to not decode its message in the second phase, denoted $\mathcal{C}_{D_1}(\Theta_1)$, is expressed as
\begin{equation}\label{eq.11}
\mathcal{C}_{D_1}(\Theta_1)\triangleq \frac{\mathrm{MRC}_{({SD_1},{RD_1})}\,a_1}{\mathrm{MRC}_{({SD_1},{RD_1})}\,a_2+I_{X_{D_1}}+I_{Y_{D_1}}} < \Theta_1,
\end{equation}
where is defined as 
\begin{equation}\label{eq.12}
 \mathrm{MRC}_{({SD_1},{RD_1})}\triangleq \vert h_{SD_1}\vert^{2}l_{SD_1}+\vert h_{RD_1}\vert^{2}l_{RD_1}
\end{equation}

In the same way, in the second phase, $D_2$ adds the power received from $S$ and from $R$. Hence, the outage event at $D_2$ to not decode $D_1$ message, denoted $\mathcal{C}_{D_{2-1}}(\Theta_1)$, and the outage event at $D_2$ to not decode its message, denoted $\mathcal{C}_{D_{2-2}}(\Theta_2)$, are respectively expressed as
\begin{equation}\label{eq.13}
\mathcal{C}_{D_{2-1}}(\Theta_1)\triangleq \frac{ \mathrm{MRC}_{({SD_2},{RD_2})}\,a_1}{ \mathrm{MRC}_{({SD_2},{RD_2})}\,a_2+I_{X_{D_2}}+I_{Y_{D_2}}} < \Theta_1,
\end{equation}
and
\begin{equation}\label{eq.14}
\mathcal{C}_{D_{2-2}}(\Theta_2)\triangleq \frac{ \mathrm{MRC}_{({SD_2},{RD_2})}\,a_2}{I_{X_{D_2}}+I_{Y_{D_2}}} < \Theta_2.
\end{equation}
The overall outage event related to $D_1$, denoted $\textit{O}_{(1)}$, is given by 
\begin{equation}\label{eq.15}
\textit{O}_{(1)}\triangleq  \Big[ \mathcal{B}_{D_1}(\Theta_1) \cap \mathcal{A}_{R_1}(\Theta_1) \Big]\cup\Big[\mathcal{A}_{R_1}^C(\Theta_1) \cap \mathcal{C}_{D_1}(\Theta_1) \Big],  
\end{equation}

Finally, the overall outage event related to $D_2$, denoted $\textit{O}_{(2)}$, is given by 
\begin{eqnarray}\label{eq.16}
\textit{O}_{(2)}&\triangleq& \left[\Bigg\{\bigcup_{i=1}^{2} \mathcal{B}_{D_{2-i}}(\Theta_i)\Bigg\} \cap \Bigg\{\bigcup_{i=1}^{2} \mathcal{A}_{R_i}(\Theta_i)\Bigg\}\right] 
\nonumber\\ &&\bigcup \left[\Bigg\{\bigcap_{i=1}^{2} \mathcal{A}_{R_{i}}^C(\Theta_i)\Bigg\} \cap \Bigg\{\bigcup_{i=1}^{2} \mathcal{C}_{D_{2-i}}(\Theta_i)\Bigg\}\right].  
\end{eqnarray}

\subsection{Outage Probability Expressions}

In the following, we will express the outage probability $\textit{O}_{(1)}$ and $\textit{O}_{(2)}$. The probability $\mathbb{P}(\textit{O}_{(1)})$, when $\Theta_1 < a_1/ a_2$, is given by

\begin{multline}\label{eq.17}
       \mathbb{P}(\textit{O}_{(1)})=1- \mathcal{J}_{(D_1)}\big(\frac{G_{1}}{l_{SD_1}}\big)-
       \mathcal{J}_{(R)}\big(\frac{G_{1}}{l_{SR}}\big)+
       \mathcal{J}_{(D_1)}\big(\frac{G_{1}}{l_{SD_1}}\big)\mathcal{J}_{(R)}\big(\frac{G_{1}}{l_{SR}}\big)\\
       +\mathcal{J}_{(R)}\big(\frac{G_{1}}{l_{SR}}\big)-
       \frac{l_{RD_1}\mathcal{J}_{(R)}\big(\frac{G_{1}}{l_{SR}}\big)\mathcal{J}_{(D_1)}\big(\frac{G_{1}}{l_{RD_1}}\big)-l_{SD_1}\mathcal{J}_{(R)}\big(\frac{G_{1}}{l_{SR}}\big)\mathcal{J}_{(D_1)}\big(\frac{G_{1}}{l_{SD_1}}\big)}{l_{RD_1}-l_{SD_1}},
\end{multline}

where $G_{1}=\Theta_1/(a_1-\Theta_1 a_2)$, and $\mathcal{J}_{(M)}\Big(\frac{A}{B}\Big)$ is expressed as  \\
\begin{equation}\label{eq.19}
    \mathcal{J}_{(M)}\Big(\frac{A}{B}\Big)=\mathcal{L}_{I_{X_{M}}}\Big(\frac{A}{B}\Big)\mathcal{L}_{I_{Y_{M}}}\Big(\frac{A}{B}\Big).
\end{equation}
The  probability $ \mathbb{P}(\textit{O}_{(2)})$, when $\Theta_1 < a_1/ a_2$, is given by 
\begin{multline}\label{eq.18}
       \mathbb{P}(\textit{O}_{(2)})=1- \mathcal{J}_{(D_2)}\big(\frac{G_{\mathrm{max}}}{l_{SD_2}}\big)-
       \mathcal{J}_{(R)}\big(\frac{G_{\mathrm{max}}}{l_{SR}}\big)+
       \mathcal{J}_{(D_2)}\big(\frac{G_{\mathrm{max}}}{l_{SD_2}}\big)\mathcal{J}_{(R)}\big(\frac{G_{\mathrm{max}}}{l_{SR}}\big)\\
       +\mathcal{J}_{(R)}\big(\frac{G_{\mathrm{max}}}{l_{SR}}\big)-
       \frac{l_{RD_2}\mathcal{J}_{(R)}\big(\frac{G_{\mathrm{max}}}{l_{SR}}\big)\mathcal{J}_{(D_2)}\big(\frac{G_{\mathrm{max}}}{l_{RD_2}}\big)-l_{SD_2}\mathcal{J}_{(R)}\big(\frac{G_{\mathrm{max}}}{l_{SR}}\big)\mathcal{J}_{(D_2)}\big(\frac{G_{\mathrm{max}}}{l_{SD_2}}\big)}{l_{RD_2}-l_{SD_2}},
\end{multline}
where $G_{\mathrm{max}}=\mathrm{max}(G_1,G_2)$, and $G_2=\Theta_2 /a_2$.\\
\textit{Proof}:  See Appendix \ref{App.A}.\hfill $ \blacksquare $ \\

\section{Laplace Transform Expressions}
 In this section, we derive the Laplace transform expressions of the interference from the $X$ road and from the $Y$ road. The Laplace transform of the interference originating from the $X$ road at the received node, denoted $M$, is expressed as

\begin{equation}\label{eq:33}
\mathcal{L}_{I_{X_{M}}}(s)=\exp\Bigg(-\emph{p}\lambda_{X}\int_\mathbb{R}\dfrac{1}{1+\Vert \textit{x}-{M} \Vert^\alpha/s}\textrm{d}x\Bigg),
\end{equation}
where 
\begin{equation}\label{eq:34}
\Vert \textit{x}-{M} \Vert=\sqrt{\big[m\sin(\theta_{{M}})\big]^2+\big[x-m \cos(\theta_{M}) \big]^2 }.
\end{equation}
The Laplace transform of the interference originating from the $Y$ road at $M$ is given by
\begin{equation}\label{eq:35}
\mathcal{L}_{I_{Y_{M}}}(s)=\exp\Bigg(-\emph{p}\lambda_{Y}\int_\mathbb{R}\frac{1}{1+\Vert \textit{y}-{M} \Vert^\alpha/s}\textrm{d}y\Bigg),
\end{equation}
where
\begin{equation}\label{eq:36}
\Vert \textit{y}-{M} \Vert=\sqrt{\big[m\cos(\theta_{{M}})\big]^2+\big[y-m \sin(\theta_{M}) \big]^2 },
\end{equation}
\textit{Proof}:  See Appendix \ref{App.B}.\hfill $ \blacksquare $ \\
The expression (\ref{eq:33}) and (\ref{eq:35}) can be calculated with mathematical
tools such as MATLAB. Closed form expressions are
obtained for $ \alpha=2 $ and  $ \alpha=4 $. We only present the expressions
when $ \alpha=2 $ due to lack of space.
 
 The Laplace transform expressions of the interference at the node ${M}$ when $\alpha=2$ are given by
\begin{equation}\label{eq:37}
\mathcal{L}_{I_{X_{M}}}(s)=\exp\Bigg(-\dfrac{\emph{p}\lambda_{X}s\pi}{\sqrt{\big[m\sin(\theta_{{M}})\big]^2+s}}\Bigg),
\end{equation}
and
\begin{equation}\label{eq:38}
\mathcal{L}_{I_{Y_{M}}}(s)=\exp\Bigg(-\dfrac{\emph{p}\lambda_{Y} s\pi}{\sqrt{\big[m\cos(\theta_{{M}})\big]^2+s}}\Bigg).
\end{equation}
\textit{Proof}:  See Appendix \ref{App.C}. \hfill$\blacksquare $

\section{Simulations and Discussions}
In this section, we evaluate the performance of cooperative NOMA using MRC at road intersections. 
In order to verify the accuracy of the theoretical results, Monte Carlo simulations are carried out by averaging over 10,000 realizations of the PPPs and fading parameters. In all figures, Monte Carlo simulations are presented by marks, and they match perfectly the theoretical results, which validates the correctness of our analysis. We set, without loss of generality,  $\lambda_X = \lambda_Y =\lambda$. Unless stated otherwise, $S=(0,0)$, $R=(50,0)$, $D_1=(100,10)$, and $D_2=(100,-10)$.

\begin{figure}[]
\centering
\includegraphics[height=8cm,width=9cm]{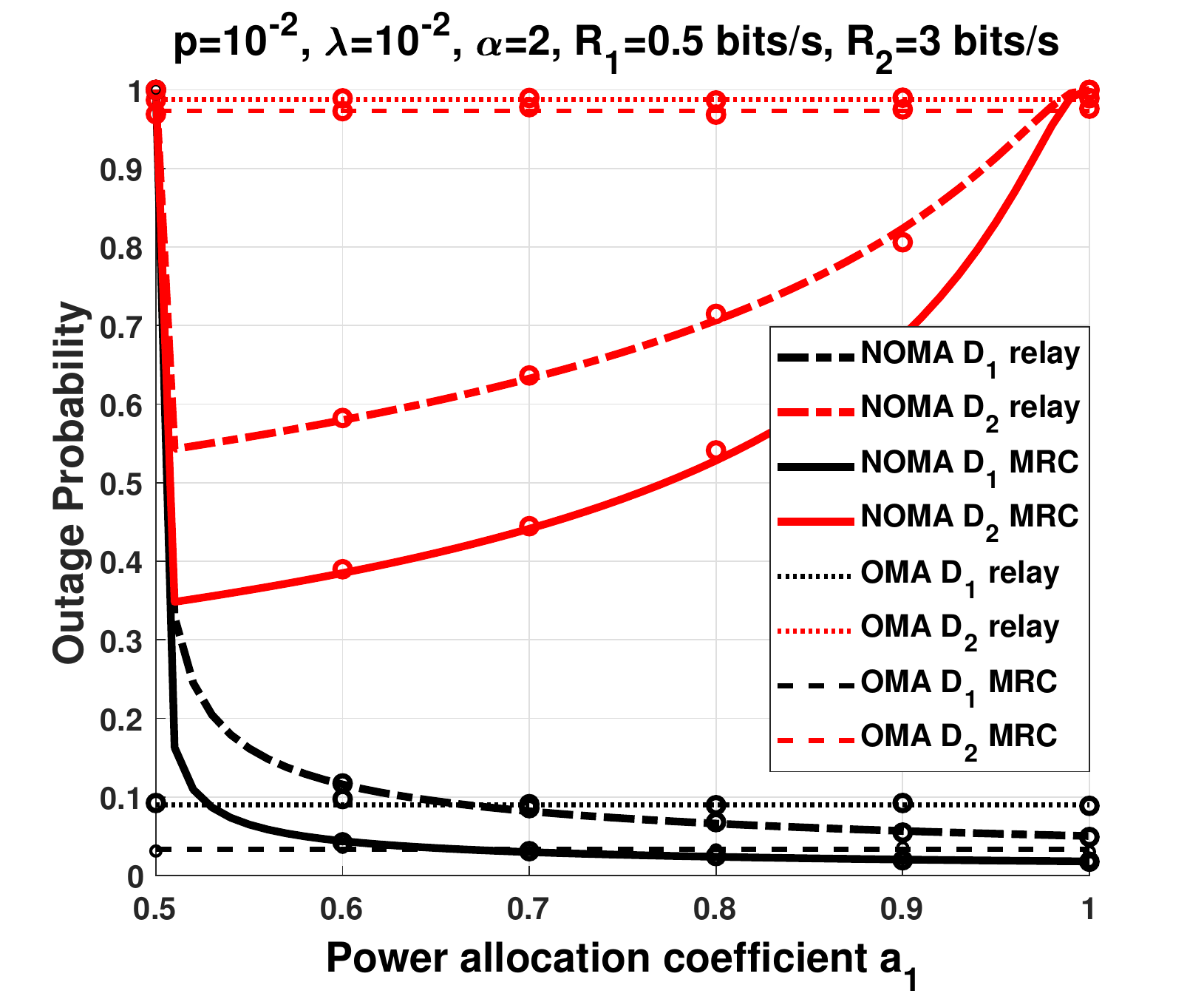}
\caption{Outage probability as a function of $a_1$, using a relay transmission and MRC transmission, considering NOMA and OMA.}
\label{Fig2}
\end{figure}

Fig.\ref{Fig2} shows the outage probability as a function of $a_1$, using a relay transmission \cite{VTC} and MRC transmission, considering NOMA and OMA. We can see from Fig.\ref{Fig2}, that using MRC offers a significant improvement over the relay transmission. We can also see that the improvement that MRC offers compared to the the relay transmission is greater for $D_2$ using NOMA. We can alos see that MRC using NOMA has a decreases in outage of $34\%$ compared to relay using NOMA. Whereas the improvement of MRC using OMA compared to relay OMA is $2\%$. On the other hand, we can notice an improve of $60\%$ when using MRC in NOMA compared to MRC in OMA.

\begin{figure}[]
\centering
\includegraphics[height=8cm,width=9cm]{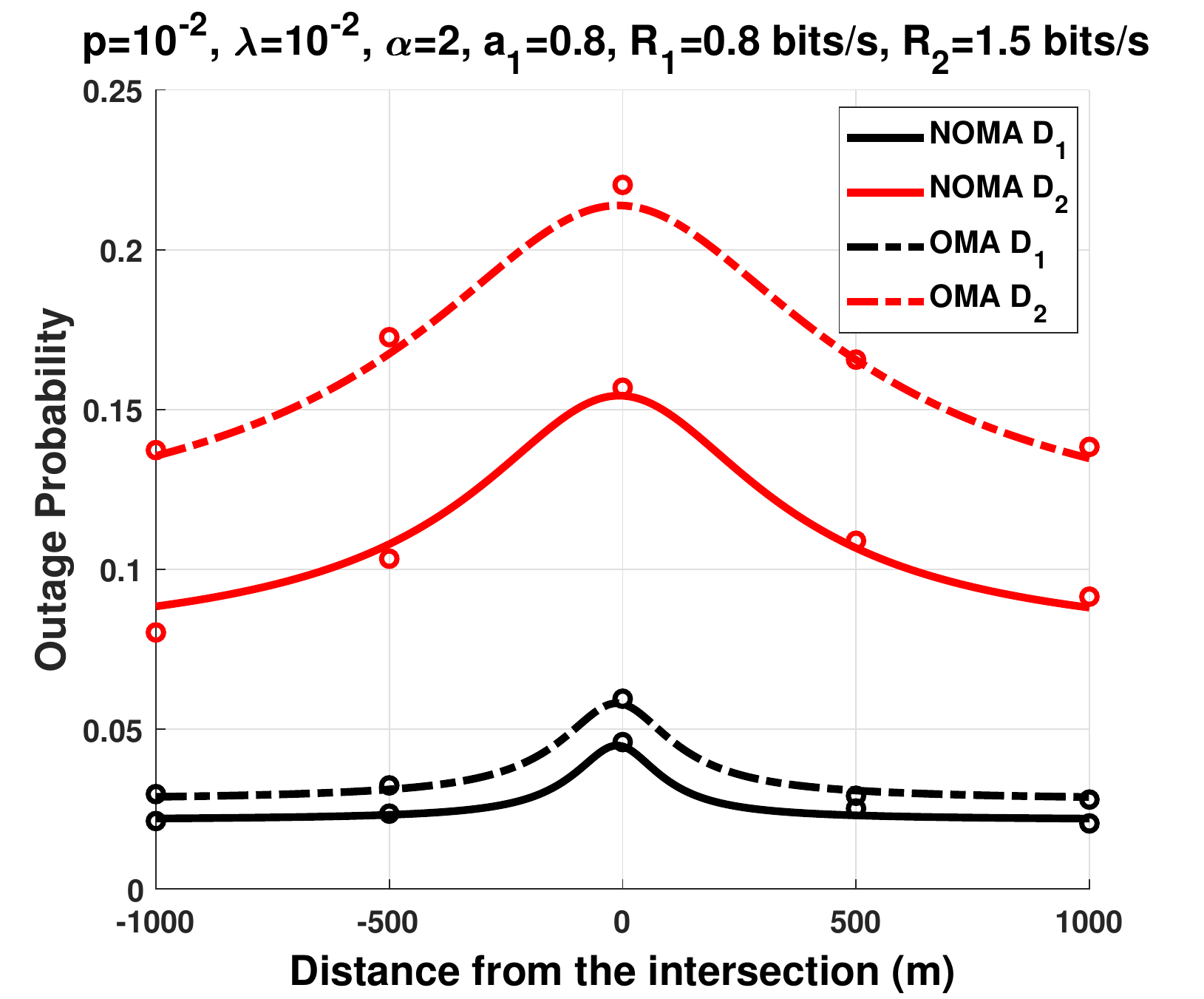}
\caption{Outage probability as a function of the distance between the nodes and the intersection, considering NOMA and OMA.}
\label{Fig3}
\end{figure}

Fig.\ref{Fig3} shows the outage probability as a function of the distance between the nodes and the intersection, considering NOMA and OMA. We can see that the outage probability reaches its maximum value a the intersection, that is, when the distance between the nodes and the intersection equals zero.
This because when the nodes are far from the intersection, the aggregate interference of the vehicles that are located on the same road as the nodes interfere is greater than the aggregate interference of the vehicles that are on the other road. 
However, when the nodes are at the intersection, the interfering vehicles of both roads interfere equally on the nodes.
We can also see from Fig.\ref{Fig3} that NOMA outperforms OMA for both $D_1$ and $D_2$. 
\begin{figure}[]
\centering
\includegraphics[height=8cm,width=9cm]{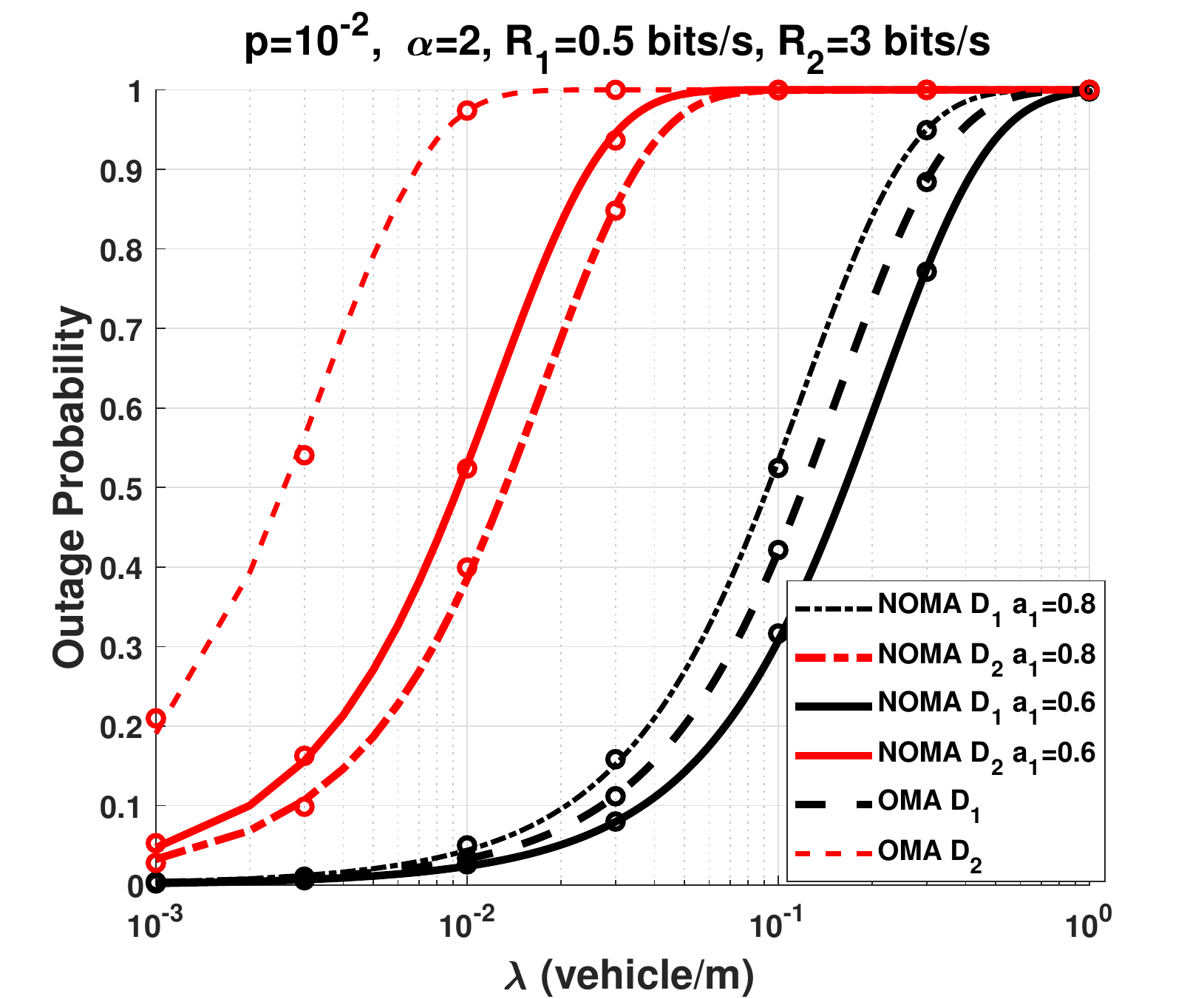}
\caption{Outage probability as a function of $\lambda$, considering NOMA and OMA. }
\label{Fig4}
\end{figure}

Fig.\ref{Fig4} investigates the impact of the vehicles density $\lambda$ on the outage probability, considering NOMA and OMA. We can see from Fig.\ref{Fig4} that, as the intensity of the vehicles increases, the outage probability increases. We can also see that, when $a_1=0.6$, NOMA outperforms OMA for both $D_1$ and $D_2$. However, we can see that, when when $a_1=0.8$, NOMA outperforms OMA only for $D_1$, whereas OMA outperforms NOMA for $D_2$. This because, when we allocate more power to $D_1$, less power is allocated to $D_2$, which decreases the performance of NOMA compared to OMA.

\begin{figure}[]
\centering
\includegraphics[height=8cm,width=9cm]{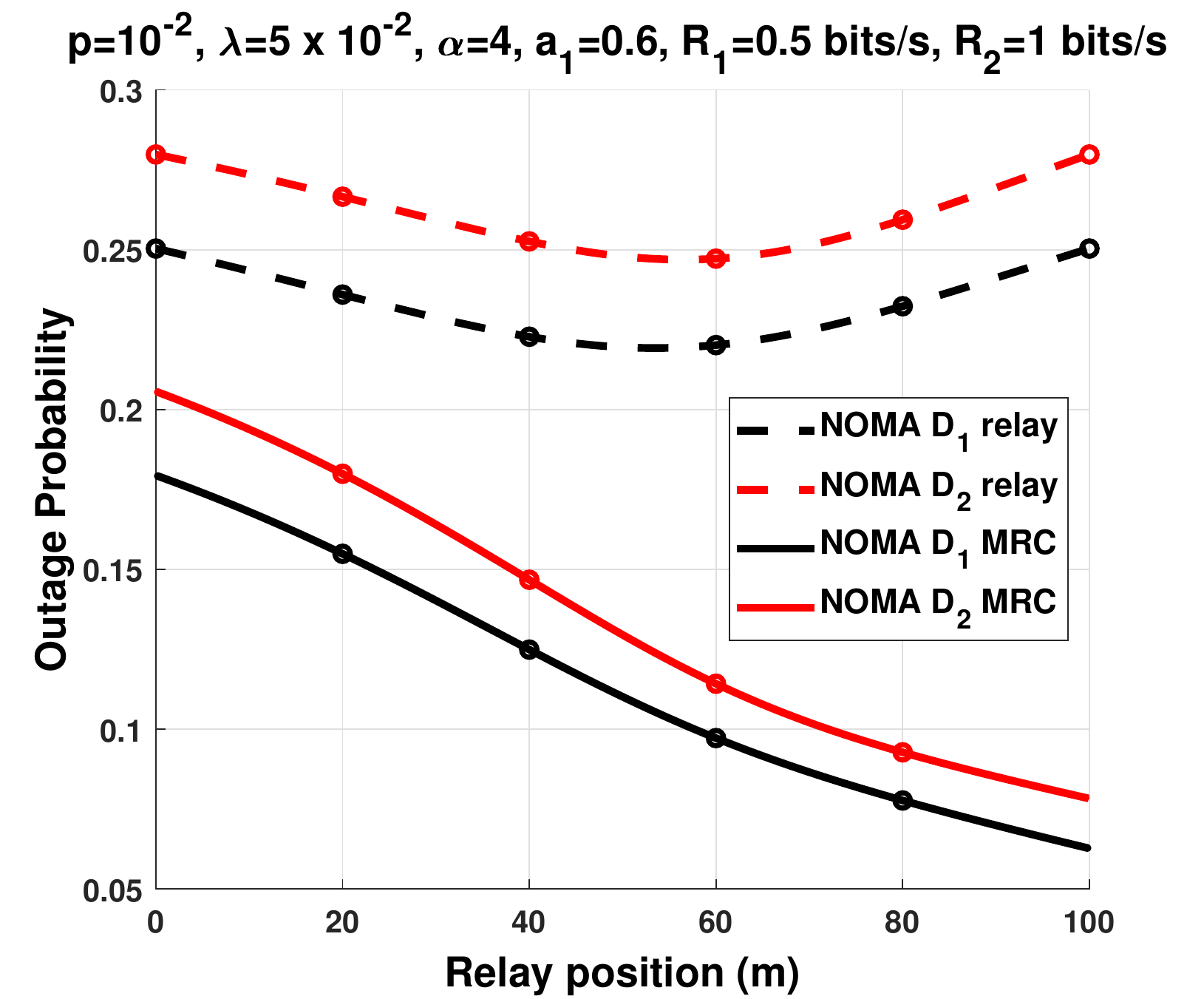}
\caption{Outage probability as a function of the relay position, using a relay transmission and MRC transmission considering NOMA.}
\label{Fig5}
\end{figure}

Fig.\ref{Fig5} depicts the outage probability as a function of the relay position, using a relay transmission and MRC transmission considering NOMA. Without loss of generality, we set $\Vert S-D_1 \Vert = \Vert S-D_2 \Vert=100$m.
We can notice from Fig.\ref{Fig5} that, the optimal position for the relay using a relay transmission is at the mid distance between the source $S$, and the destinations, $D_1$ and $D_2$. However, we can see that for MRC, the optimal relay position is when the relay is close to the destination nodes. This can be explained as follows: when the relay is close to the destination ($D_1$ or $D_2$), the channel between $S$ and $D_1$ ($S \rightarrow D_1$) and the channel between $R$ and $D_1$ ($R \rightarrow D_1$) will be decorrelated, thus, increasing the diversity gain. 

\section{Conclusion}
In this paper, we studied the improvement of using MRC in cooperative VCs transmission schemes considering NOMA at intersections. Closed form outage probability expressions were obtained.
We compared the performance of MRC cooperative NOMA with a classical cooperative NOMA, and showed that MRC in cooperative NOMA transmission offers a significant improvement over the classical cooperative NOMA in terms of outage probability. We also compared the performance of MRC cooperative NOMA with MRC cooperative orthogonal multiple access (OMA), and we showed that NOMA has a better performance than OMA. Finally, we showed that the outage probability increases when the nodes come closer to the intersection, and that using MRC considering NOMA improves the performance in this context.

\appendices
\section{}\label{App.A}

The outage probability related to $D_1$, denoted $\mathbb{P}(\textit{O}_{(1)})$, is expressed as
\begin{equation}\label{eq.26}
\mathbb{P}(\textit{O}_{(1)})=\mathbb{P}\Big( \mathcal{B}_{D_1} \cap \mathcal{A}_{D_1} \Big)+\mathbb{P}\Big(\mathcal{A}_{D_1}^C \cap \mathcal{C}_{D_1} \Big) 
\end{equation}
First, we calculate the probability $\mathbb{P}\Big(\mathcal{A}_{R_1}^C \cap \mathcal{C}_{D_1} \Big)$ as 
\begin{eqnarray}\label{eq.20}
\mathbb{P}\Big(\mathcal{A}_{R_1}^C \cap \mathcal{C}_{D_1} \Big)&=& \mathbb{E}_{I_{X},I_Y}\Bigg[\mathbb{P}\Bigg\lbrace\ \frac{\vert h_{SR}\vert^{2}l_{SR}a_1}{\vert h_{SR}\vert^{2}l_{SR}a_2+I_{X_{R}}+I_{Y_{R}}} \ge \Theta_1\nonumber\\
&& \bigcap \frac{\left(\vert h_{SD_1}\vert^{2}l_{SD_1}+\vert h_{RD_1}\vert^{2}l_{RD_1}\right)\,a_1}{\left(\vert h_{SD_1}\vert^{2}l_{SD_1}+\vert h_{RD_1}\vert^{2}l_{RD_1}\right)\,a_2+I_{X_{D_1}}+I_{Y_{D_1}}} < \Theta_1\Bigg\rbrace\Bigg]  \\
&=& \mathbb{E}_{I_{X},I_Y}\Bigg[\mathbb{P}\Bigg\lbrace\ \vert h_{SR}\vert^{2}l_{SR}(a_1-\Theta_1 a_2)\ge\Theta_1\big[I_{X_{R}}+I_{Y_{R}}\big]\nonumber\\
&&\bigcap\left(\vert h_{SD_1}\vert^{2}l_{SD_1}+\vert h_{RD_1}\vert^{2}l_{RD_1}\right)\,(a_1-\Theta_1 a_2) < \Theta_1\big[I_{X_{D_1}}+I_{Y_{D_1}}\big]\Bigg\rbrace\Bigg]. \nonumber\\
\end{eqnarray}
When $\Theta_1 < a_1/ a_2$, and setting $G_{1}= \Theta_1 /(a_1- \Theta_1 a_2)$, we obtain    
\begin{eqnarray}\label{eq.30}
\mathbb{P}\Big(\mathcal{A}_{R_1}^C \cap \mathcal{C}_{D_1} \Big)&=& \mathbb{E}_{I_{X},I_Y}\Bigg[\mathbb{P}\Bigg\lbrace\ \vert h_{SR}\vert^{2}\ge \frac{G_1}{l_{SR}}\big[I_{X_{R}}+I_{Y_{R}}\big]\Bigg\rbrace
\nonumber\\&&\times \Bigg\lbrace 1-\mathbb{P}\left(\vert h_{SD_1}\vert^{2}l_{SD_1}+\vert h_{RD_1}\vert^{2}l_{RD_1} \ge G_1\big[I_{X_{D_1}}+I_{Y_{D_1}}\big]\right)\Bigg\rbrace\Bigg]. 
\end{eqnarray}
Since $|h_{SR}|^2$ follows an exponential distribution with unit mean, and the second probability in (\ref{eq.30}) can be written as
\begin{multline} \label{eq.21}
\mathbb{P}\big[\vert h_{SD_1}\vert^{2}l_{SD_1}+\vert h_{RD_1}\vert^{2}l_{RD_1} \ge G_1 (I_{X_{D_1}}+I_{Y_{D_1}})\big]=\\
\dfrac{l_{RD_1} \exp\Big[- \dfrac{G_1}{l_{RD_1}}(I_{X_{D_1}}+I_{Y_{D_1}})\Big]- l_{SD_1}\exp\Big(- \dfrac{G_1}{l_{SD_1}}[I_{X_{D_1}}+I_{Y_{D_1}})\Big]}{l_{RD_1}-l_{SD_1}}.
\end{multline}

Then, the equation (\ref{eq.30}) becomes
\begin{multline}\label{eq.22}
\mathbb{P}\Big(\mathcal{A}_{R_1}^C \cap \mathcal{C}_{D_1} \Big)= \mathbb{E}_{I_{X},I_Y}\Bigg[ \exp\left( \frac{G_1}{l_{SR}}\big[I_{X_{R}}+I_{Y_{R}}\big]\right)\\ \times  1-\dfrac{l_{RD_1} \exp\Big[- \dfrac{G_1}{l_{RD_1}}(I_{X_{D_1}}+I_{Y_{D_1}})\Big]- l_{SD_1}\exp\Big(- \dfrac{G_1}{l_{SD_1}}[I_{X_{D_1}}+I_{Y_{D_1}})\Big]}{l_{RD_1}-l_{SD_1}}\big]\Bigg\rbrace\Bigg]  \\
= \mathbb{E}_{I_{X},I_Y}\Bigg[ \exp\left( \frac{G_1}{l_{SR}}\big[I_{X_{R}}+I_{Y_{R}}\big]\right) -  \exp\left( \frac{G_1}{l_{SR}}\big[I_{X_{R}}+I_{Y_{R}}\big]\right)  \\
\times \dfrac{l_{RD_1} \exp\Big[- \dfrac{G_1}{l_{RD_1}}(I_{X_{D_1}}+I_{Y_{D_1}})\Big]- l_{SD_1}\exp\Big(- \dfrac{G_1}{l_{SD_1}}[I_{X_{D_1}}+I_{Y_{D_1}})\Big]}{l_{RD_1}-l_{SD_1}}\big]\Bigg\rbrace\Bigg] .
\end{multline}
Using the independence of the PPP on the road $X$ and $Y$, and given that  $\mathbb{E}[e^{sI}]=\mathcal{L}_I(s)$,  we finally get
\begin{multline}\label{eq.23}
\mathbb{P}\Big(\mathcal{A}_{R_1}^C \cap \mathcal{C}_{D_1} \Big)= \mathcal{L}_{I_{X_{R}}}\bigg(\frac{G_{1}}{l_{SR}}\bigg)\mathcal{L}_{I_{Y_{R}}}\bigg(\frac{G_{1}}{l_{SR}}\bigg)\\-\mathcal{L}_{I_{X_{R}}}\bigg(\frac{G_{1}}{l_{SR}}\bigg)\mathcal{L}_{I_{Y_{R}}}\bigg(\frac{G_{1}}{l_{SR}}\bigg)\frac{l_{RD_1}\mathcal{L}_{I_{X_{D_1}}}\bigg(\frac{G_{1}}{l_{RD_1}}\bigg)\mathcal{L}_{I_{Y_{D_1}}}\bigg(\frac{G_{1}}{l_{RD_1}}\bigg)-l_{SD_1}\mathcal{L}_{I_{X_{D_1}}}\bigg(\frac{G_{1}}{l_{SD_1}}\bigg)\mathcal{L}_{I_{Y_{D_1}}}\bigg(\frac{G_{1}}{l_{SD_1}}\bigg)}{l_{RD_1}-l_{SD_1}}.
\end{multline}

The probability $\mathbb{P}\Big( \mathcal{B}_{D_1} \cap \mathcal{A}_{D_1} \Big)$ can be expressed as 
\begin{eqnarray}\label{eq.32}
\mathbb{P}\Big( \mathcal{B}_{D_1} \cap \mathcal{A}_{D_1} \Big)&=&1-\mathbb{P}\Big( \mathcal{B}_{D_1}^C \cup \mathcal{A}_{D_1}^C \Big)\nonumber\\&=&1-\mathbb{P}\Big(\mathcal{B}_{D_1}^C\Big)-\mathbb{P}\Big(\mathcal{A}_{D_1}^C\Big)+\mathbb{P}\Big(\mathcal{B}_{D_1}^C \cap \mathcal{A}_{D_1}^C\Big)
\end{eqnarray}
The final expression can acquired following the same steps above.

The outage probability related to $D_2$, denoted $\mathbb{P}(\textit{O}_{(2)})$, is expressed as

\begin{eqnarray}\label{eq.34}
\mathbb{P}(\textit{O}_{(2)})&=& \mathbb{P}\left[\Bigg\{\bigcup_{i=1}^{2} \mathcal{B}_{D_{2-i}}(\Theta_i)\Bigg\}\cap\Bigg\{\ \bigcup_{i=1}^{2} \mathcal{A}_{R_i}(\Theta_i)\Bigg\}\right] 
\nonumber\\&& + \mathbb{P}\left[\Bigg\{\bigcap_{i=1}^{2} \mathcal{A}_{R_{i}}^C(\Theta_i)\Bigg\} \cap \Bigg\{ \bigcup_{i=1}^{2} \mathcal{C}_{D_{2-i}}(\Theta_i)\Bigg\}\right].  
\end{eqnarray}
To calculate the first probability in (\ref{eq.34}), we proceed as follows
\begin{eqnarray}\label{eq.35}
\mathbb{P}\left[\Bigg\{\bigcup_{i=1}^{2} \mathcal{B}_{D_{2-i}}(\Theta_i)\Bigg\}\cap \Bigg\{\bigcup_{i=1}^{2} \mathcal{A}_{R_i}(\Theta_i)\Bigg\}\right]\nonumber&=&1- \mathbb{P}\left[\Bigg\{\bigcap_{i=1}^{2} \mathcal{B}_{D_{2-i}}^C(\Theta_i)\Bigg\}\cup \Bigg\{\bigcap_{i=1}^{2} \mathcal{A}_{R_i}^C(\Theta_i)\Bigg\}\right]\nonumber\\
&=&1- \mathbb{P}\left[\bigcap_{i=1}^{2} \mathcal{B}_{D_{2-i}}^C(\Theta_i)\right]- \mathbb{P}\left[\bigcap_{i=1}^{2} \mathcal{A}_{R_{i}}^C(\Theta_i)\right]\nonumber\\
&&+\mathbb{P}\left[\Bigg\{\bigcap_{i=1}^{2} \mathcal{B}_{D_{2-i}}^C(\Theta_i)\Bigg\}\cap \Bigg\{\bigcap_{i=1}^{2} \mathcal{A}_{R_i}^C(\Theta_i)\Bigg\}\right].\nonumber\\
\end{eqnarray}
Since the computation of first and the second probability in (\ref{eq.35}) follow the same steps above, we only  calculate the last probability in (\ref{eq.35}), hence, proceed as follows 
\begin{multline}\label{eq.24}
\mathbb{P}\left[\Bigg\{\bigcap_{i=1}^{2} \mathcal{B}_{D_{2-i}}^C(\Theta_i)\Bigg\}\cap \Bigg\{\bigcap_{i=1}^{2} \mathcal{A}_{R_i}^C(\Theta_i)\Bigg\}\right] =\\
\mathbb{E}_{I_{X},I_Y}\Bigg[\mathbb{P}\Bigg\lbrace\frac{\vert h_{SD_2}\vert^{2}l_{SD_2}a_1}{\vert h_{SD_2}\vert^{2}l_{SD_2}a_2+I_{X_{D_2}}+I_{Y_{D_2}}} \ge \Theta_1, 
\frac{\vert h_{SD_2}\vert^{2}l_{SD_2}a_2}{I_{X_{D_2}}+I_{Y_{D_2}}} \ge \Theta_2,\\ \frac{\vert h_{SR}\vert^{2}l_{SR}a_1}{\vert h_{SR}\vert^{2}l_{SR}a_2+I_{X_{R}}+I_{Y_{R}}} \ge \Theta_1, 
\frac{\vert h_{SR}\vert^{2}l_{SR}a_2}{I_{X_{R}}+I_{Y_{R}}} \ge \Theta_2\Bigg\rbrace\Bigg].
\end{multline}

When $\Theta_1 < a_1/ a_2$, and setting $G_{2}= \theta_2 /a_2$, we obtain
\begin{multline}\label{eq.25}
\mathbb{P}\left[\Bigg\{\bigcap_{i=1}^{2} \mathcal{B}_{D_{2-i}}^C(\Theta_i)\Bigg\}\cap \Bigg\{\bigcap_{i=1}^{2} \mathcal{A}_{R_i}^C(\Theta_i)\Bigg\}\right] =\\
\mathbb{E}_{I_{X},I_Y}\Bigg[\mathbb{P}\Bigg\lbrace\vert h_{SD_2}\vert^{2}\ge \frac{G_1}{l_{SD_2}}\big[I_{X_{D_2}}+I_{Y_{D_2}}\big], 
\vert h_{SD_2}\vert^{2}\ge \frac{G_2}{l_{SD_2}}\big[I_{X_{D_2}}+I_{Y_{D_2}}\big],\\ \vert h_{SR}\vert^{2}\ge \frac{G_1}{l_{SR}}\big[I_{X_{R}}+I_{Y_{R}}\big], 
\vert h_{SR}\vert^{2}\ge \frac{G_2}{l_{SR}}\big[I_{X_{R}}+I_{Y_{R}}\big]\Bigg\rbrace\Bigg]\\
=\mathbb{E}_{I_{X},I_Y}\Bigg[\mathbb{P}\Bigg\lbrace\vert h_{SD_2}\vert^{2}\ge \frac{\max(G_1,G_2)}{l_{SD_2}}\big[I_{X_{D_2}}+I_{Y_{D_2}}\big], \vert h_{SR}\vert^{2}\ge \frac{\max(G_1,G_2)}{l_{SR}}\big[I_{X_{R}}+I_{Y_{R}}\big]\Bigg\rbrace\Bigg].
\end{multline}
Finally, we get
\begin{eqnarray}\label{eq.36}
\mathbb{P}\left[\Bigg\{\bigcap_{i=1}^{2} \mathcal{B}_{D_{2-i}}^C(\Theta_i)\Bigg\}\cap \Bigg\{\bigcap_{i=1}^{2} \mathcal{A}_{R_i}^C(\Theta_i)\Bigg\}\right]&=&
\mathcal{L}_{I_{X_{D_2}}}\bigg(\frac{G_{\mathrm{max}}}{l_{SD_2}}\bigg)\mathcal{L}_{I_{Y_{D_2}}}\bigg(\frac{G_{\mathrm{max}}}{l_{SD_2}}\bigg)\nonumber\\
&&\times\:\mathcal{L}_{I_{X_{R}}}\bigg(\frac{G_{\mathrm{max}}}{l_{SR}}\bigg)\mathcal{L}_{I_{Y_{R}}}\bigg(\frac{G_{\mathrm{max}}}{l_{SR}}\bigg),
\end{eqnarray}
where $G_{\mathrm{max}}=\mathrm{max}(G_1,G_2)$.

\section{}\label{App.B}
The Laplace transform of the interference originating from the X road at $M$ is expressed as 
\begin{equation}\label{eq:63}
\mathcal{L}_{{I_{X_{M}}}}(s)=\mathbb{E}\big[{\\\exp(-sI_{X_{M}})}\big].
\end{equation}
Plugging (\ref{eq.5}) into (\ref{eq:63}) yields
\begin{eqnarray}\label{eq:64}
\mathcal{L}_{{I_{X_{M}}}}(s)&=&\mathbb{E}\Bigg[{\exp\Bigg(-\sum_{x\in\Phi_{X_{M}}}s\vert h_{{M}x}\vert^2 l_{{M}x}  \Bigg)}\Bigg] \nonumber\\
&=& \mathbb{E}\Bigg[\prod_{x\in\Phi_{X_{M}}} \exp\Bigg(-s\vert h_{{M}x}\vert^2l_{{M}x}\Bigg)\Bigg]\nonumber\\
&\overset{(a)}{=}&\mathbb{E}\Bigg[\prod_{x\in\Phi_{X_{M}}}\mathbb{E}_{\vert  h_{{M}x}\vert^2, p}\Bigg\lbrace \exp\Bigg(-s\vert h_{{M}x}\vert^2l_{Mx}\Bigg)\Bigg\rbrace\Bigg]\nonumber\\
&\overset{(b)}{=}&\mathbb{E}\Bigg[\prod_{x\in\Phi_{X_{M}}}\dfrac{p}{1+s l_{{M}x}}+1-p\Bigg]\nonumber \\
&\overset{(c)}{=}&\exp\Bigg(-\lambda_{X}\displaystyle\int_{\mathbb{R}}\Bigg[1-\bigg(\dfrac{p}{1+sl_{{M}x}}+1-p\bigg)\Bigg]\textrm{d}x\Bigg)\nonumber \\
&=&\exp\Bigg(-p\lambda_{X}\displaystyle\int_{\mathbb{R}}\dfrac{1}{1+1/sl_{{M}x}}\textrm{d}x\Bigg), 
\end{eqnarray}
where (a) follows from the independence of the fading coefficients; (b) follows from performing the expectation over $|h_{{M}x}|^2$ which follows an exponential distribution with unit mean, and performing the expectation over the set of interferes; (c) follows from the probability generating functional (PGFL) of a PPP \cite{haenggi2012stochastic}. 
Then, substituting $l_{{M}x}=\Vert \textit{x}-M \Vert^{-\alpha}$ in (\ref{eq:64}) yields (\ref{eq:33}). The equation (\ref{eq:35}) can be acquired by following the same steps.

\section{} \label{App.C}

In order to calculate the Laplace transform of interference originated from the $X$ road at the node $M$, we have to calculate the integral in (\ref{eq:33}). We calculate the integral in (\ref{eq:33}) when $\alpha=2$. Let us take $m_{x}=m \cos(\theta_{M})$, and $m_{y}=m \sin(\theta_{M}$), then (\ref{eq:33}) becomes
\begin{eqnarray}\label{eq:65}
\mathcal{L}_{I_{X_{M}}}(s)&=&\exp\Bigg(-\emph{p}\lambda_{X}\int_\mathbb{R}\dfrac{1}{1+m_{y}^2+(x-m_{x})^2/s }\textrm{d}x\Bigg)\nonumber\\
&=&\exp\Bigg(-\emph{p}\lambda_{X}s\int_\mathbb{R}\dfrac{1}{s+m_{y}^2+(x-m_{x})^2 }\textrm{d}x\Bigg).\nonumber\\
\end{eqnarray}
The integral inside the exponential in (\ref{eq:65}) equals
\begin{equation}\label{eq:66}
\int_\mathbb{R}\dfrac{1}{s+m_{y}^2+(x-m_{x})^2}\textrm{d}x=\dfrac{\pi}{\sqrt{m_{y}^2+s}}.
\end{equation}
Then, plugging (\ref{eq:66}) into (\ref{eq:65}), we obtain
\begin{equation}\label{eq:67}
\mathcal{L}_{I_{X_M}}(s)=\exp\Bigg(-\emph{p}\lambda_{X}\dfrac{s\,\pi}{\sqrt{m_{y}^2+s}}\Bigg).
\end{equation}
Finally, substituting $m_{y}$ by $m\sin(\theta_{{M}})$ into (\ref{eq:67}) yields (\ref{eq:37}). Following the same steps above, and without details for the derivation, we obtain (\ref{eq:38}).
\bibliographystyle{ieeetr}
\bibliography{bibnoma}

\end{document}